\documentclass{aa}     

\usepackage{graphicx}
\usepackage{txfonts}
\usepackage{natbib}
\usepackage{url}

\bibpunct{(}{)}{;}{a}{}{,}

\newcommand{\Mpc}{$h^{-1}$\thinspace Mpc}

\def\apjl{ApJL}
\def\apjs{ApJS}

\newcommand{\vmh}{h^{-1}\mathrm{Mpc} }

\begin{document}

\title{SDSS DR7 superclusters} 
\subtitle{Principal component analysis}

\author { M.  Einasto\inst{1} \and L.J. Liivam\"agi\inst{1,2}  \and E.  Saar\inst{1,3} 
\and J.  Einasto\inst{1,3,4} \and  E.  Tempel\inst{1}  \and E.  Tago\inst{1} 
\and V.J. Mart\'{\i}nez\inst{5}
}

\institute{Tartu Observatory, 61602 T\~oravere, Estonia
\and 
Institute of Physics, Tartu University, T\"ahe 4, 51010 Tartu, Estonia
\and
Estonian Academy of Sciences,  EE-10130 Tallinn, Estonia
\and
ICRANet, Piazza della Repubblica 10, 65122 Pescara, Italy\and 
Observatori Astron\`omic, Universitat de Val\`encia, Apartat
de Correus 22085, E-46071 Val\`encia, Spain
}

\authorrunning{M. Einasto et al. }

\offprints{M. Einasto}

\date{ Received   ... / Accepted  ...   }

\titlerunning{PCA}

\abstract
{
The study of  superclusters of galaxies helps us to understand
the formation, evolution, and present-day properties of the large-scale 
structure of the Universe.
}
{We use data about superclusters  drawn from the SDSS DR7
to analyse  possible  selection effects in the
supercluster catalogue, to study the 
physical and morphological properties of superclusters,
to find their possible subsets, and to determine  scaling relations
for our superclusters.
 }
{We apply principal component analysis and Spearman's correlation
test to study the properties of superclusters.
   }
{
We have found that the  parameters of superclusters do not correlate with their distance. 
The correlations between the physical and 
morphological properties of superclusters are strong. 
Superclusters can be divided into two 
populations according to their total luminosity:
high-luminosity ones with 
$L_g > 400$ $10^{10}h^{-2} L_{\sun}$, and low-luminosity systems. 
High-luminosity superclusters form two sets, which are
more elongated systems with the 
shape parameter $K_1/K_2 < 0.5$ and less elongated ones with
$K_1/K_2 > 0.5$. 
The first two 
principal components account for more than 90\% 
of the variance in the supercluster 
parameters. We use principal component analysis to derive scaling relations 
for superclusters, in which we combine the physical and morphological
parameters of superclusters. 
}
{The first two principal
components define the fundamental plane,  which
 characterises  the physical and morphological properties of superclusters.
Structure formation simulations for different cosmologies, and 
more data about the local and high redshift superclusters
are needed to understand the evolution and the properties of superclusters better.
}


\maketitle

\section{Introduction} 
\label{sect:intro} 
The large-scale distribution of the 
dark and baryonic matter in the Universe can be described as the cosmic web --
the network of galaxies, groups, and clusters of galaxies connected by filaments 
\citep{1978MNRAS.185..357J, 1978ApJ...222..784G, zes82, 1986ApJ...302L...1D}. In 
this network superclusters  are the largest density enhancements 
formed by the density perturbations on a scale of about 100~\Mpc\ 
($H_0=100 h \mathrm{km\,s^{-1}Mpc^{-1}})$.
Numerical simulations show that high-density peaks in the density
distribution (the seeds of supercluster cores) are seen already at
very early stages of the formation and evolution of structure
\citep{2010AIPC.1205...72E}.  These are the locations of the formation
of the first objects in the Universe
\citep[e.g.][]{2004A&A...424L..17V,2005ApJ...635..832M,
  2005ApJ...620L...1O, 2011MNRAS.410.1537H}. 
Studying the properties of superclusters helps us to understand
the formation, evolution, and  properties of the large-scale 
structure of the Universe \citep[][and 
references therein]{2007JCAP...10..016H, 2009MNRAS.399...97A, 
2010MNRAS.409..156B}.
Comparison of  observed and simulated superclusters, especially extreme systems
among them, is a test of cosmological
models \citep{kbp02, 2007A&A...462..397E, e07, 2009MNRAS.399...97A,
2011ApJ...736...51E, 2011arXiv1105.3378S}.

The first step in supercluster studies is to compile  
supercluster catalogues, which serve as observational databases. Supercluster catalogues have been constructed using the
friend-of-friend method or using a smoothed density field of galaxies. The first method 
has been applied to the data on rich (Abell) clusters of galaxies to obtain 
catalogues of superclusters of rich clusters, both from
observations and simulations \citep{1993ApJ...407..470Z, 
1994MNRAS.269..301E, 1995A&AS..113..451K, 1997A&AS..123..119E,
e2001, 2006ApJ...652..907W}. 
Density field 
superclusters have been determined using data of deep surveys of galaxies 
\citep{bas03,2003A&A...405..425E, 2004MNRAS.352..939E, 2006A&A...459L...1E, 
2007A&A...462..811E, 2010arXiv1012.1989J, 2011MNRAS.411.1716C,  
2011MNRAS.415..964L}. The properties of 
superclusters have been studied, for example, 
by  \citet{1998A&A...336...35J}, \citet{kbp02}, \citet{2011MNRAS.411.1716C}, 
\citet{2011MNRAS.415..964L}, \citet{2006ApJ...652..907W}, and
\citet{e2001, 2007A&A...462..397E, 2007A&A...464..815E, e07, 
2011A&A...532A...5E}. 
These studies show that the properties of superclusters are correlated. More 
luminous superclusters are richer and larger, contain richer galaxy clusters,
and have higher maximum densities of galaxies
than less luminous systems. 
High-luminosity superclusters are more elongated and 
have more complicated inner structure than 
low-luminosity ones.

In the present paper we   use the Spearman's correlation test and the principal 
component analysis (PCA), an excellent tool for multivariate data analysis, to 
investigate how strong  the correlations between the properties of superclusters 
are.  Our goals are to analyse the presence of possible distance-dependent 
selection effects in the supercluster catalogue, to study the correlations 
between the physical and morphological properties of superclusters,  to find the
possible subsets and outliers of superclusters, and to determine the scaling 
relations for the superclusters.

Principal component analysis have been used in astronomy  for a number of 
purposes:  the 
study of the properties of stars \citep{1964PTarO..34..156T, 1964MNRAS.127..493D}, 
spectral  classification of galaxies \citep[][and references 
therein]{2010ApJ...714..487S}, morphological classification of galaxies 
\citep{2010arXiv1009.0723C}, studies of galaxies, galaxy groups,  and 
dark matter haloes \citep[][and references therein] {1984MNRAS.206..453E, 
2004ApJ...600..640L, 2006MNRAS.370..828F, 2008MNRAS.390.1453W, 
2010arXiv1009.0030C, 2011A&A...527A..49I, 2011ApJ...732...93T, 
2011arXiv1103.1641S, 2011MNRAS.415L..69J},
for the Hubble parameter reconstruction 
\citep[][and references therein]{2011A&A...527A..49I},
and for studies of star formation history in the universe using
gamma ray bursts \citep{2011arXiv1106.1745I}.
Our study is the first in which the 
PCA is applied to explore the properties of superclusters of galaxies. 

In Sect.~\ref{sect:data} we give data about superclusters. 
In Sect.~\ref{sect:pca} we describe the PCA 
and the Spearman's  correlation test, and apply them in Sect.~\ref{sect:results} 
to study the physical and morphological properties of superclusters and
to derive scaling relations 
for the superclusters. We discuss selection effects in Sect.~\ref{sect:sel}
and give our conclusions in Sect.~\ref{sect:disc}.

We assume  the standard cosmological parameters: the Hubble parameter $H_0=100~ 
h$ km s$^{-1}$ Mpc$^{-1}$, the matter density $\Omega_{\rm m} = 0.27$, and the 
dark energy density $\Omega_{\Lambda} = 0.73$.

\section{Data}
\label{sect:data}

\begin{figure*}[ht]
\centering
\resizebox{0.96\textwidth}{!}{\includegraphics*{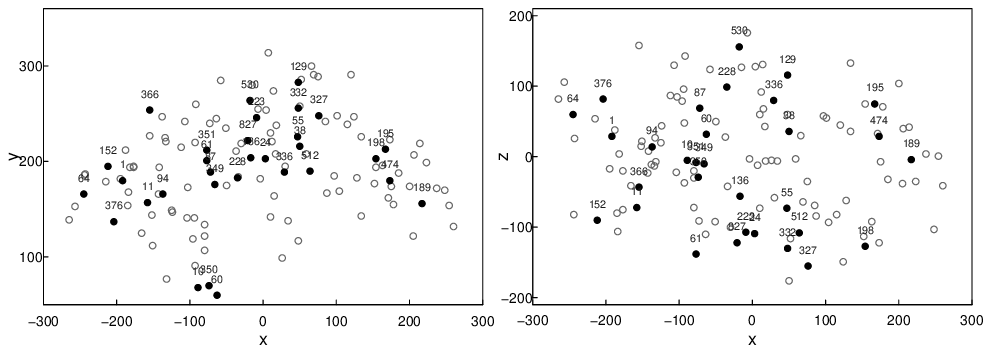}}
\caption{The distribution of  superclusters in cartesian coordinates, 
in units of \Mpc. 
The filled circles denote superclusters with the luminosity 
$L_g > 400$ $10^{10}h^{-2} L_{\sun}$,
empty circles denote less luminous superclusters.
The numbers are ID's of luminous supercluster  from L10 (Table~\ref{tab:basicscl}).
}
\label{fig:sclxyz}
\end{figure*}

We selected the MAIN galaxy sample of the 7th data release of the Sloan
Digital Sky Survey \citep{ade08,abaz08}
with the apparent $r$ magnitudes $12.5 \leq r \leq 17.77$,   excluding
duplicate entries. The sample  is described in detail in
\citet{2010A&A...514A.102T}, hereafter T10. 
We corrected the redshifts
of galaxies for the motion relative to the CMB and computed the co-moving
distances \citep{mar03} of galaxies.

We calculated the galaxy luminosity density field to reconstruct the underlying mass distribution. 
To determine superclusters (extended systems of galaxies) in the luminosity 
density field we created a set of density contours by choosing a density 
threshold and define connected volumes above a certain density threshold as 
superclusters. In order to choose proper density levels to determine 
individual superclusters, we analysed the  density field superclusters at a 
series of density levels. As a result we used  the density level $D = 5.0$ (in 
units of mean density;
mean luminosity density of our sample is 
$\ell_{\mathrm{mean}}$ = 1.526$\cdot10^{-2}$ $\frac{10^{10} h^{-2} L_\odot}{(\vmh)^3})$
 to determine  individual superclusters. At this density 
level superclusters in the richest chains of superclusters in the volume under 
study  still form separate systems; at lower density levels they join into huge 
percolating systems. At higher threshold density levels superclusters are 
smaller and their number decreases. 

In our flux-limited catalogue the luminosity-dependent selection effects are the 
smallest at the distance interval 90~\Mpc\ $\le D_\mathrm{com} \le $ 320~\Mpc. 
For the present study we chose superclusters of galaxies in this 
distance interval. There are 125 superclusters in the sample. 
Even the poorest systems in our sample contain several groups of 
galaxies. These systems can be compared with the Local supercluster
containing one cluster of galaxies with outgoing filaments.
In the 
Appendix~\ref{sec:DF} we give the details of the calculations of galaxy 
luminosities  and  of the luminosity density field, as well as of the 
selection effects. The description of the supercluster catalogues is given in 
\citet[][hereafter L10]{2010arXiv1012.1989J}. 
\footnote{
The supercluster catalogues can be downloaded from: 
\url{http://atmos.physic.ut.ee/~juhan/super/}. 
}

The superclusters can be characterised by the following  physical  parameters: the total weighted luminosity of galaxies in a
supercluster, $L_g$, the volume $Volume$, the diameter 
$Diameter$, and the number of 
galaxies in superclusters, $N_{\mbox{gal}}$.
The supercluster volume  is calculated 
from the density field as the number of connected grid cells multiplied by the 
cell volume:
\begin{equation}
    Volume = N_{scl}\Delta^3,
    \label{eq:vol}
\end{equation}
where $\Delta$ is the grid cell length. 

The total luminosity of the superclusters $L_g$ is calculated 
as  the sum of weighted galaxy luminosities:
\begin{equation}
    L_g = \sum_{\mathrm{gal} \in \mathrm{scl}} W_L (d_{\mathrm{gal}}) L_{\mathrm{gal}}.
    \label{eq:wlum}
\end{equation}
Here the $W_L(d_{\mathrm{gal}})$ 
is the distance-dependent weight of a galaxy (the ratio of the 
expected total luminosity to the luminosity within the visibility window). We 
describe the calculation of weights in Appendix~\ref{sec:DF}. The diameter of a 
supercluster is defined as  the maximum distance between its galaxies.  
The distance of a supercluster is the distance to it's density 
maximum. 
The peak density $D_{\mbox{peak}}$ is that of the highest 
density peak within the supercluster. Usually the highest values of densities 
coincide with the richest cluster of galaxies in a supercluster.
For details we refer to L10. 

The overall morphology of a supercluster is described by the  shapefinders 
$K_1$ (planarity) and $K_2$ (filamentarity), and their ratio, $K_1$/$K_2$ (the 
shape parameter). 
The shapefinders are calculated using the volume, area, and
integrated mean 
curvature of a supercluster; they contain information both about the 
sizes of superclusters and about their outer shape. 
Systems with different shapes and similar sizes have different
shape parameters \citep{e08}. 
For the first time the shapefinders were applied  in the studies of
galaxy systems by \citet{2001MNRAS.323...47B} who analysed the shapes
of the PSCz superclusters.
We use the maximum value of 
the fourth Minkowski functional $V_3$ (the clumpiness) to characterise 
the inner structure of the superclusters. The larger the value of $V_3$, the 
more complicated the inner morphology of a supercluster is; superclusters may be 
clumpy, and they also may have holes or tunnels in them \citep{e07, 
2011ApJ...736...51E}.The formulae for the Minkowski functionals and 
shapefinders are given in App.\ref{sec:MF}.

The large-scale distribution of   superclusters is shown in
Fig.~\ref{fig:sclxyz}  in cartesian coordinates.
These coordinates are defined as in
\citet{2007ApJ...658..898P} and in \citet{2010arXiv1012.1989J}:
\begin{equation}
\begin{array}{l}
    x = -d \sin\lambda, \nonumber\\[3pt]
    y = d \cos\lambda \cos \eta,\\[3pt]
    z = d \cos\lambda \sin \eta,\nonumber
\end{array}
\label{eq:xyz}
\end{equation}
where $d$ is the comoving distance, and $\lambda$ and $\eta$ are the SDSS 
survey coordinates. \citet{2011A&A...532A...5E} gave detailed description of the 
large-scale distribution of rich superclusters.

\section{Principal component analysis}
\label{sect:pca}

\begin{figure*}[ht]
\centering
\resizebox{0.90\textwidth}{!}{\includegraphics*{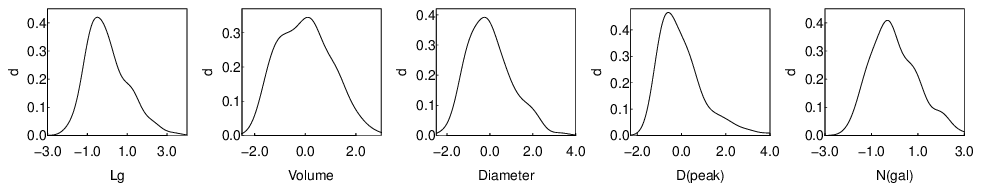}}
\caption{Distribution of the standardised physical parameters of superclusters.
From left to right: the total weighted luminosity of galaxies
$L_g$,   the volume and the diameter of superclusters, the density of the highest 
density peak inside superclusters, $D_{\mbox{peak}}$, and the number of 
galaxies in superclusters, $N_{\mbox{gal}}$. 
}
\label{fig:scl90para5distr}
\end{figure*}

The idea of the principal component analysis is to find a small number of 
linear combinations of correlated parameters  to describe most of the 
variation in the dataset with a small number of new uncorrelated parameters. The PCA 
transforms the data to a new coordinate system, where the greatest variance by 
any projection of the data lies along the first coordinate (the first principal 
component), the second greatest variance -- along the second coordinate, and so on. 
There are as many principal components as there are parameters, but typically 
only the first few are needed to explain most of the total variation. 

Principal components PC$x$  ($x \in \mathbb{N}$, $x \leq N_{tot})$ 
are a linear combination of the original parameters:

\begin{equation}\label{eq:pc}
 PCx = \sum_{i=1}^{N_{\mathrm{tot}}} a(i)_{x} V_{i}
\end{equation}
where $-1 \leq a(i)_x \leq 1$ are the coefficients of the linear transformation,
$V_i$ are the original parameters and $N_{\mathrm{tot}}$ is the number of the original 
parameters.

PCA is suitable tool to study simultaneously correlations between
a large number of parameters, for finding subsets in data, and  detecting outliers. 
Linear combinations of principal 
components can be used to reproduce parameters characterising objects in the 
dataset. 

Principal components can be used to derive scaling relations. If data points lie 
along a plane, defined by the first two principal components, then the scaling 
relations along this plane are defined by the third principal component
\citep{1984MNRAS.206..453E}.
For the  analysis we use standardised parameters, centred on their means 
($ V_{i} - \overline{V_{i}})$ and normalised (divided by their standard deviations,
$\sigma( V_{i}))$. 
Therefore we obtain for the scaling relations:

\begin{equation}\label{eq:scaling}
 \sum_{i=1}^{N_{\mathrm{tot}}} a(i)_{3} \frac{(V_{i} - \overline{V_{i}})}{\sigma (V_{i})} = 0.
\end{equation}

\begin{figure}[ht]
\centering
\resizebox{0.48\textwidth}{!}{\includegraphics*{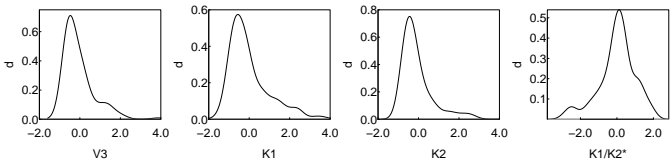}}
\caption{Distribution of the standardised morphological parameters of superclusters.
From left to right: 
the maximum value of the fourth Minkowski functional $V_3$, 
the planarity $K_1$, the filamentarity $K_2$, and the shape parameter of superclusters,
$K_1/K_2^*$.
}
\label{fig:scl90para4distr}
\end{figure}

For PCA, the parameters should be normally distributed.
Therefore we use the logarithms of parameters in most cases; this makes the distributions
more gaussian, and the range over which their values span are smaller, especially
for luminosities and volumes. We do not use logarithms of morphological data,
in order to not to exclude from the analysis those with negative
values of shapefinders, which may occur in the case of
compact superclusters with a complex overall morphology 
\citep{e08, 2011ApJ...736...51E}.
Figures~\ref{fig:scl90para5distr} and \ref{fig:scl90para4distr} 
show the distribution of the values of the standardised parameters. Deviations from
the normal distribution are mostly caused by the most luminous (or by the poorest
for the shape parameter) 
superclusters in our sample.  
In Table~\ref{tab:msd} we give the mean values and standard deviations
of supercluster parameters. For poor superclusters
of ``spider'' morphology the shape parameter is not always well defined
\citep{2011A&A...532A...5E}. For five systems 
the value of the shape parameter $|K_1/K_2| > 4$; therefore we also calculated
the mean value and standard deviation of the shape parameter without
these systems (denoted as $K_1/K_2^*$). This effect does not affect the values of
other parameters, thus we did not exclude these systems from our
calculations.

We present in tables  the values of principal components and  the standard 
deviations, proportion of variance, and cumulative variance of principal 
components. The values of components show the importance of the original parameters 
in each PCx. We plot the principal planes for superclusters.
For the calculations we used command 
{\it prcomp} from {\it R}, an open-source free statistical environment developed 
under the GNU GPL \citep[][\texttt{http://www.r-project.org}]{ig96}.

To study correlations between  properties of superclusters,
we applied Spearman's rank correlation test, in which the value of the
correlation coefficient $r$ shows the presence of correlation 
($r = 1$ for perfect correlation), anticorrelation ($r = -1$
for perfect anticorrelation),
or the absence of correlations when $r \approx 0$.  

\begin{table}[ht]
\caption{Mean values and standard deviations of supercluster parameters.}
\begin{tabular}{lrr} 
\hline\hline 
(1)&(2)&(3)\\      
\hline 
Parameter       & $mean$ & $sd$  \\
\hline
$\log(L_g)$        & 2.367 & 0.378   \\
$\log(Volume)$     & 2.813 & 0.571   \\
$\log(Diameter)$   & 1.179 & 0.258   \\
$\log(D_{\mbox{peak}})$   & 0.856 & 0.119   \\
$\log(N_{\mbox{gal}})$    & 2.219 & 0.435   \\
$\log(Dist.)$      & 2.379 & 0.113   \\
 $V_3$          & 1.770  & 1.185   \\
 $K_1$          & 0.015  & 0.031   \\
 $K_2$          & 0.027  & 0.069  \\
 $K_1/K_2$      & -0.050 & 3.701   \\
 $K_1/K_2^*$      & 0.338 & 0.756   \\
\hline            
                
\label{tab:msd}  
\end{tabular}\\
\tablefoot{ 
$L_g$ -- 
the total weighted luminosity of galaxies in 
superclusters in units of $10^{10}h^{-2} L_{\sun}$;   
$Volume$ -- in units of $(\vmh)^{3}$; 
$Diameter$ -- in Mpc/$h$; 
$N_{\mbox{gal}}$ -- the number of 
galaxies in superclusters; $D_{\mbox{peak}}$ -- the density of the highest 
density peak inside superclusters, in units of mean density;  
$Dist$ -- the distance in Mpc/$h$; $V_3$ is 
the maximum value of the fourth Minkowski functional, 
$K_1$ is the planarity, $K_2$ is the filamentarity, and the ratio, $K_1$/$K_2$,
is the shape parameter of superclusters (see Section~\ref{sect:data}
for definitions). $K_1/K_2^*$ denotes the shape parameter for the supercluster
sample from which we excluded five most noisy values as explained in the text.
}
\end{table}

\section{Results}
\label{sect:results}

\subsection{PCA with physical parameters of superclusters}
\label{subsect:pcaall}

We start the calculations of principal components  using physical 
characteristics of superclusters and  their distances. 
Including the supercluster
distances may show possible correlations between the other parameters 
of superclusters and their distance, which will indicate that the parameters of 
superclusters are affected by distance-dependent selection effects.

\begin{table}[ht]
\caption{Results of the principal component analysis,
with the distances of superclusters included.}
\begin{tabular}{lrrr} 
\hline\hline  
(1)&(2)&(3)&(4)\\      
\hline
               &    PC1   &   PC2   &   PC3     \\
$\log(N_{\mbox{gal}})$      &   -0.444 &   0.264 &  -0.108    \\  
$\log(L_g)$       &   -0.455 &  -0.149 &  -0.097    \\  
$\log(Diameter)$  &   -0.441 &  -0.133 &  -0.542    \\  
$\log(Volume)$      &   -0.454 &  -0.126 &  -0.042    \\  
$\log(D_{\mbox{peak}})$  &   -0.427 &  -0.062 &   0.825    \\  
$\log(Distance)$  &    0.100 &  -0.932 &   0.012    \\   

\hline
\multicolumn{3}{l}{Importance of components} & \\ 
\hline
     &          PC1  &   PC2  &   PC3     \\
Standard deviation             &  2.148  & 1.046  & 0.466  \\
Proportion of Variance   &  0.769  & 0.182  & 0.036  \\
Cumulative Proportion    &  0.769  & 0.951  & 0.987  \\
\hline

\label{tab:pca90dist}  
\end{tabular}\\
\tablefoot{ 
Notations  given in Section~\ref{sect:data}.
}
\end{table}

\begin{table}[ht]
\caption{Results of the Spearman's rank correlation test.}
\begin{tabular}{lrr} 
\hline\hline 
(1)&(2)&(3)\\      
\hline 
Parameters     & $r$ & $p$  \\
\hline
$\log(Dist.)$ vs. $\log(L_g)$         & -0.06  & 0.50     \\
$\log(Dist.)$ vs. $\log(N_{\mbox{gal}})$     & -0.49  & $9.8e-9$  \\
$\log(Dist.)$ vs. $\log(Diameter)$       & -0.11  & 0.20      \\
$\log(Dist.)$ vs. $\log(Volume)$        & -0.08  & 0.40    \\
$\log(Dist.)$ vs. $\log(D_{\mbox{peak}})$    & -0.09  & 0.33    \\
$\log(Dist.)$ vs. $V_3$            & -0.03  & 0.78  \\
$\log(Dist.)$ vs. $K_1$            & -0.08  & 0.37      \\
$\log(Dist.)$ vs. $K_2$            &  0.04  & 0.70    \\
$\log(Dist.)$ vs. $K_1/K_2$        & -0.05  & 0.58    \\
\\
$\log(L_g)$ vs. $\log(N_{\mbox{gal}})$    & 0.88  & $< 2.2e-16$   \\
$\log(L_g)$ vs. $\log(Diameter)$      & 0.95  & $< 2.2e-16$   \\
$\log(L_g)$ vs. $\log(Volume)$       & 0.98  & $< 2.2e-16$   \\
$\log(L_g)$ vs. $\log(D_{\mbox{peak}})$   & 0.94  & $< 2.2e-16$   \\
\\
$\log(L_g)$ vs. $V_3$     & 0.75  & $< 2.2e-16$   \\
$\log(L_g)$ vs. $K_1$     & 0.89  & $< 2.2e-16$   \\
$\log(L_g)$ vs. $K_2$     & 0.82  & $< 2.2e-16$   \\
$\log(L_g)$ vs. $K_1/K_2$ & 0.19  & 0.04          \\
\hline            
                
\label{tab:rank}  
\end{tabular}\\
\tablefoot{ 
Rank correlation coefficient  $r$ and the p-value $p$.
The values $p < 0.05$ mean that the results are statistically of very
high significance.
}
\end{table}

Table~\ref{tab:pca90dist}  presents the results of this analysis. We show the 
values of only the first three principal components, enough for this 
test.  The coefficients of the first principal component of the physical 
parameters  are of almost equal value, while the coefficient 
corresponding to the distance is very small -- the first principal component 
accounts for most of the variance of the physical parameters of superclusters. 
The second principal component accounts for most of the variance of the 
distances of superclusters.  This shows that the physical parameters of 
superclusters are not correlated with distance. To ensure 
that this interpretation is correct we carried out the Spearman's tests for 
correlations (Table~\ref{tab:rank}). These tests showed a weak anticorrelation 
between the distance  and the number of galaxies in 
superclusters, with a high statistical significance. This is not 
surprising since the catalogue of superclusters is based on the flux-limited 
sample  in which the number of galaxies in superclusters depends on 
the distance. The sample of superclusters was chosen from a relatively narrow 
distance interval, so this dependence is weak. For other parameters of 
superclusters (luminosity, diameter, volume, and peak density), the tests showed 
a very weak correlation with distance (Spearman's rank $r \approx 0.1$ or 
less), but with no statistical significance, as the $p$-values show. Therefore we 
conclude that there are no  correlations between the distances and physical 
parameters of superclusters, and the distance-dependent selection effects have been 
 properly taken into account when generating the supercluster catalogue and 
calculating the physical properties of superclusters.

\begin{table}[ht]
\caption{Results of the principal component analysis for the physical
parameters.
}
\begin{tabular}{lrrrrr} 
\hline\hline 
(1)&(2)&(3)&(4)&(5)&(6)\\      
\hline 
            &   PC1   &   PC2   &   PC3   &  PC4 &  PC5     \\
\hline
$\log(N_{\mbox{gal}})$  &   -0.439 &   0.056 &   0.895 &  -0.036 &  -0.018    \\
$\log(L_g)$      &   -0.460 &   0.112 &  -0.217 &  -0.047 &   0.851   \\
$\log(Diameter)$ &   -0.445 &   0.557 &  -0.238 &   0.561 &  -0.344   \\
$\log(Volume)$     &   -0.458 &   0.058 &  -0.268 &  -0.761 &  -0.367   \\
$\log(D_{\mbox{peak}})$ &   -0.430 &  -0.818 &  -0.149 &   0.319 &  -0.144   \\
\hline
\multicolumn{3}{l}{Importance of components} && \\ 
\hline
     &          PC1  &   PC2  &   PC3  &  PC4 &  PC5    \\
St. deviation    & 2.139  & 0.467 &  0.377 &  0.193 &  0.161   \\
Prop. Variance   & 0.915  & 0.043 &  0.028 &  0.007 &  0.005   \\
Cum. Proportion  & 0.915  & 0.958 &  0.987 &  0.994 &  1.000   \\
\hline

\label{tab:pca90}  
\end{tabular}\\
\tablefoot{ 
Notations as in Table~\ref{tab:pca90dist}.
}
\end{table}

We will proceed with the analysis of superclusters,
taking only the physical parameters into account. 
Table~\ref{tab:pca90}, which presents the results of this analysis,
demonstrates that the coefficients of the first principal
component are almost equal for different  parameters of superclusters.
 Therefore the parameters,
which describe the full supercluster (the luminosity, richness, 
diameter, and volume), are almost equally
important in determining the supercluster properties. 
The cumulative variance in Table~\ref{tab:pca90} shows that the first two principal 
components account for more than 95\% of the total variance in this supercluster 
sample. The first principal component accounts for most of the variance of 
the overall parameters of superclusters. The values of the second principal 
component show that the largest remaining variance in the sample comes from the peak 
density of superclusters. The values of the third principal component show
that the coefficients corresponding to the luminosity, volume, and diameter
have almost equal negative values, while the number of
galaxies has large positive coefficients. 

The PCA therefore  suggests that the physical parameters of superclusters 
are strongly correlated.
We checked for the presence of the correlations between the 
parameters with   Spearman's tests, which
showed that the correlations between the parameters of superclusters 
are statistically of very high significance, 
both between the overall parameters of superclusters and between the
overall parameters and the peak density inside the superclusters
 (Table~\ref{tab:rank}). We only present the correlations between
 the luminosity and other parameters, to keep Table~\ref{tab:rank}
 short. The results of the tests of other correlations are similar.
Especially tight are the correlations between the luminosities, the
diameters, and the volumes of superclusters, as the correlation 
coefficients show in Table~\ref{tab:rank}.

\begin{figure}[ht]
\centering
\resizebox{0.48\textwidth}{!}{\includegraphics*{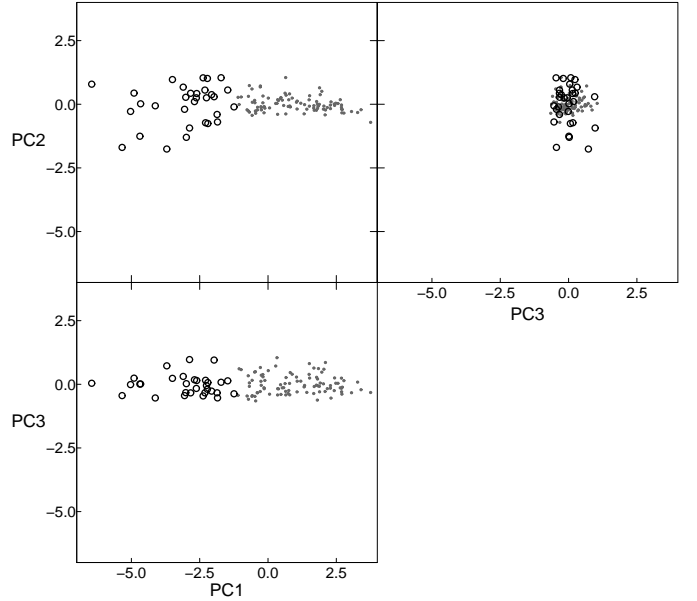}}
\caption{Principal planes for superclusters, PCA with physical parameters.
Open circles: high-luminosity superclusters with 
luminosity $L_g > 400$ $10^{10}h^{-2} L_{\sun}$,
grey dots: superclusters of lower luminosity.
}
\label{fig:scl90pc12}
\end{figure}

Let us take a look at the locations of superclusters 
in the principal planes (Fig.~\ref{fig:scl90pc12}). 
The upper lefthand panel shows the distribution of superclusters in the
principal plane PC1-PC2. Most superclusters form here an elongated 
cloud with a very small scatter. 
 These are
low-luminosity superclusters with the luminosity $L_g < 400$ 
$10^{10}h^{-2} L_{\sun}$.
The scatter of positions of high-luminosity superclusters  is larger. 
This suggests that we can divide superclusters into two populations
according to their total luminosity. The transition between populations is smooth.
We give the data about high-luminosity superclusters in 
Table~\ref{tab:basicscl}.
The luminous superclusters  
with a high value of the peak density have higher 
negative values for the second PC, and the supercluster
SCl~001 has the largest negative value of PC2. The superclusters 
with a lower value of the peak density have positive values 
of the second PC. The richest supercluster in the sample,
SCl~061, is among them. This supercluster has the highest negative value of
PC1. 
The lefthand panels of Figure~\ref{fig:scl90pc12} show that
  the more luminous the supercluster, the higher is the 
negative value of it's first principal component. The value of the peak density 
inside superclusters determines the location of superclusters along the axis of 
the second principal component.
In PC1-PC3 plane (lower left panel of Fig.~\ref{fig:scl90pc12}) superclusters
also form an elongated cloud with larger scatter of high-luminosity superclusters.
Upper right panel (PC3-PC2 plane) shows the third view of this cloud.
Such an elongated,  prolate shape is characteristic of the planar
distribution on PC1-PC2 plane \citep{2008MNRAS.390.1453W}, 
which defines the fundamental plane for superclusters.

\subsection{PCA for the  morphological parameters of superclusters}
\label{subsect:pcamorf}

\begin{table}[ht]
\caption{Results of principal component analysis for the 
luminosity and morphological properties of superclusters.}
\begin{tabular}{lrrrrr} 
\hline\hline 
(1)&(2)&(3)&(4)&(5)& (6)\\      
\hline 
     &   PC1  &   PC2  &   PC3  &  PC4   &   PC5  \\
\hline
$\log(L_g)$ &  -0.489 &   0.004 & -0.655 &   0.373 &  -0.437  \\
$V_3$      &  -0.490 &  -0.044 &  0.596 &   0.608 &   0.173  \\
$K_1$      &  -0.511 &  -0.023 & -0.297 &  -0.331 &   0.734  \\
$K_2$      &  -0.505 &  -0.056 &  0.351 &  -0.615 &  -0.488  \\
$K1/K2$    &  -0.059 &   0.997 &  0.042 &  -0.016 &  -0.000  \\
\hline
\multicolumn{3}{l}{Importance of components} &&& \\ 
\hline
     &          PC1  &   PC2  &   PC3  &  PC4   &   PC5  \\
St.deviation    & 1.891  & 0.996  & 0.528 &  0.313 &  0.235  \\
Prop.Variance   & 0.715  & 0.198  & 0.055 &  0.019 &  0.011  \\
Cum.Proportion  & 0.715  & 0.913  & 0.969 &  0.988 &  1.000  \\
\hline

\label{tab:pcamorf}  
\end{tabular}\\
\tablefoot{ 
As in Section~\ref{sect:data}.
} 
\end{table}

Next, we use the PCA to study the morphological and physical 
properties of superclusters simultaneously.   
From the physical characteristics we only include the total luminosity, which 
is sufficient since the physical parameters of superclusters are strongly 
correlated.
Table~\ref{tab:rank} shows that the mophological  parameters of 
superclusters are not 
correlated with their distances. 
Table~\ref{tab:pcamorf} shows 
 the results of PCA  for the  luminosity and the morphological parameters. 
Here the absolute values of  components for the luminosity,
the clumpiness, and the shapefinders $K_1$ and $K_2$ are almost 
equal.  Therefore the luminosity 
and these morphological parameters are equally important in 
shaping the properties of superclusters. 
The second principal component accounts for most of the variance of the
shape parameter $K_1/K_2$. 
The higher the negative value of the PC1 for the supercluster, the more
luminous  the supercluster, has higher value planarities and
filamentarities, and higher 
maximal value  of the fourth Minkowski functional $V_3$, hence a
richer inner morphology.  

Table~\ref{tab:pcamorf}  shows that the first two principal components account 
for about 93\% of the total variance in the data. 

The  Spearman's tests (Table~\ref{tab:rank}) showed 
that the correlations between the supercluster luminosity and its
morphological parameters 
are statistically  highly significant.
The correlation between the luminosity and the shape parameter of superclusters
is weak.

\begin{figure}[ht]
\centering
\resizebox{0.48\textwidth}{!}{\includegraphics*{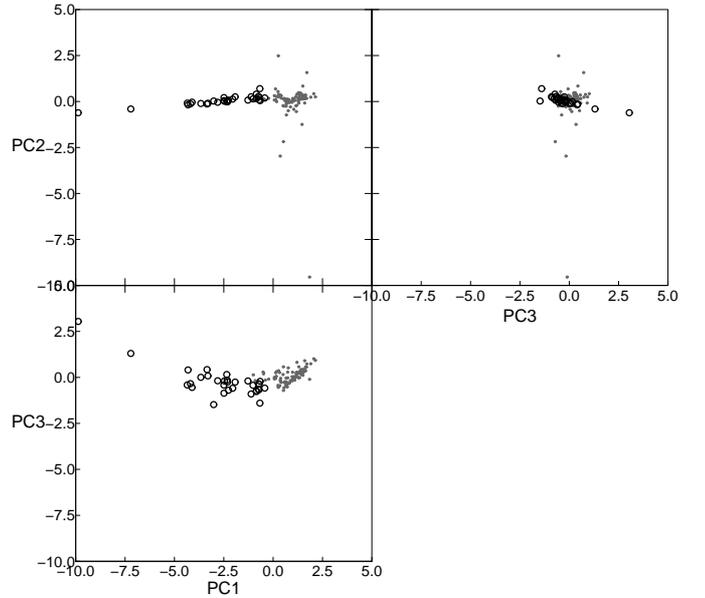}}
\caption{Principal planes for superclusters. PCA with the
morphological parameters.
Open circles: high-luminosity superclusters with 
luminosity $L_g > 400$ $10^{10}h^{-2} L_{\sun}$;
grey dots: superclusters of lower luminosity.
}
\label{fig:scl90mpc12}
\end{figure}

Figure~\ref{fig:scl90mpc12}  presents the locations of superclusters 
in the principal planes, defined by the luminosity and morphological
parameters of superclusters. 
The upper lefthand panel shows the distribution of superclusters in the
principal plane PC1-PC2. Here both high- and low-luminosity superclusters
form an elongated cloud with  very small scatter.
The scatter of positions of the high-luminosity superclusters in PC1-PC3
plane is greater. 
Again, the more luminous the supercluster, the higher  the 
negative value of its first principal component. High values of PC1 
(and the highest values of PC3) 
correspond to luminous superclusters with high values of clumpiness $V_3$
(Table~\ref{tab:pcamorf}). Large scatter along the second principal
component PC2 in principal planes correspond to superclusters with
high values of the shape parameter $K_1/K_2$. These are poor superclusters
of ``spider'' morphology, for which the shape parameter is not well defined
\citep{2011A&A...532A...5E}.
We see that the luminosity and  the morphological
parameters of superclusters  also define a  fundamental plane
for superclusters, where the physical and morphological properties 
are combined.

\subsection{Scaling relations for superclusters
and the fundamental plane
}
\label{sect:fund}

\begin{table}[ht]
\caption{Results of principal component analysis for
luminosity, diameters, and shapefinders.}
\begin{tabular}{lrrr} 
\hline\hline 
(1)&(2)&(3)&(4)\\      
\hline 
     &   PC1  &   PC2  &   PC3   \\
\hline
$\log(L_g)$    & -0.5713 &   0.7905 &  -0.2205    \\ 
  $K1D$      & -0.5833 &  -0.2020 &   0.7867    \\ 
  $K2D$      & -0.5773 &  -0.5781 &  -0.5765    \\ 
\hline
 & \multicolumn{3}{l}{Importance of components}  \\ 
\hline
     &          PC1  &   PC2  &   PC3    \\
St.deviation    & 1.696  & 0.308 &  0.165    \\
Prop.Variance   & 0.959  & 0.031 &  0.009    \\
Cum.Proportion  & 0.959  & 0.990 &  1.000    \\
\hline

\label{tab:pca3scale}  
\end{tabular}\\
\tablefoot{ 
$\log(L_g)$: logarithm of the total luminosity
of superclusters, $K1D = (1 - K_1) \cdot \log(Diameter)$, and 
$K2D = (1 - K_2) \cdot \log(Diameter)$.
} 
\end{table}

\begin{figure}[ht]
\centering
\resizebox{0.48\textwidth}{!}{\includegraphics*{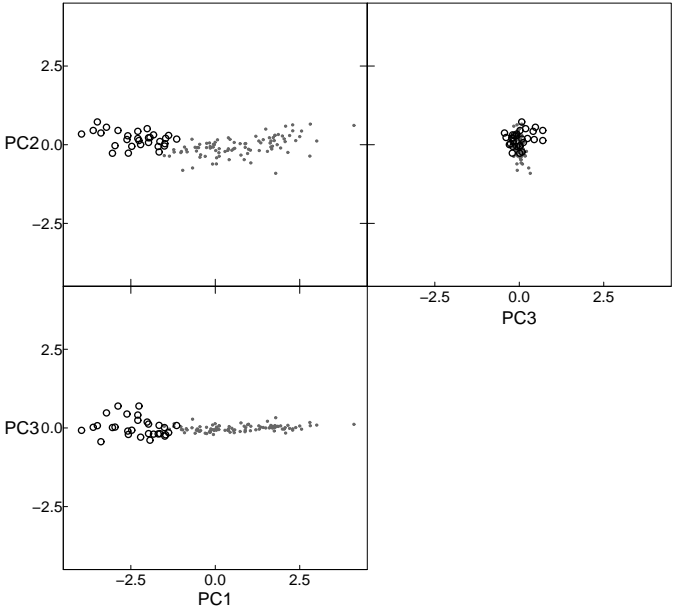}}
\caption{Principal planes for superclusters. PCA for the  
luminosity, diameter, and shapefinders as described in the text.
Open circles: high-luminosity superclusters with 
luminosity $L_g > 400$ $10^{10}h^{-2} L_{\sun}$,
grey dots: superclusters of lower luminosity.
}
\label{fig:scl90scale3pc123}
\end{figure}

The results of the PCA suggest that the first two principal
components define the fundamental plane for superclusters. This motivates us 
to find the scaling relations between the supercluster parameters.
The  scaling relations have earlier been found
between the properties of 
galaxies, of  groups of galaxies and of dark
matter haloes \citep[][and references therein]{1976ApJ...204..668F,
1977A&A....54..661T, 1977ApJ...218..333K, 1984MNRAS.206..453E, 
1987ApJ...313...59D,
1987ApJ...313...42D, 1993MNRAS.263L..21S, 1998A&A...331..493A, 2004ApJ...600..640L, 
2008ApJ...685..875D, 2008MNRAS.390.1453W, 2009MNRAS.400.1317A}.

For scaling relations we use Eq.~(\ref{eq:scaling})
and  perform the PCA for the parameters
$\log(L_g)$, $(1 - K_1)\cdot\log(Diameter)$  and $(1 - K_2)\cdot\log(Diameter)$.
This set combines the easily detectable diameter of superclusters, and
morphological parameters $K_1$ and $K_2$, 
which characterise the sizes and the shapes  of superclusters, with
the total luminosity of superclusters. 
For low values of shapefinders,
$(1 - K_1)$ and $(1 - K_2)$ are less noisy than $K_1$ and $K_2$
\citep{2011A&A...532A...5E}.

\begin{figure}[ht]
\centering
\resizebox{0.40\textwidth}{!}{\includegraphics*{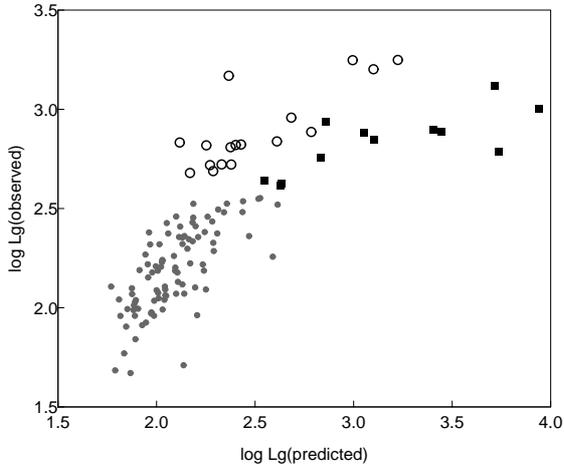}}
\caption{
$L_{g{\mathrm{(observed)}}}$ vs. $L_{g{\mathrm{(predicted)}}}$,
in units of $10^{10}h^{-2} L_{\sun}$.
Open circles denote high-luminosity superclusters with 
the luminosity $L_g >  400$ $10^{10}h^{-2} L_{\sun}$ and the
shape parameter $K_1/K_2 > 0.5$
(less elongated superclusters), filled squares denote
high-luminosity superclusters with the shape parameter $K_1/K_2 < 0.5$
(more elongated superclusters),
grey dots denote superclusters of lower luminosity.
}
\label{fig:scl90scale}
\end{figure}

Table~\ref{tab:pca3scale} and
Fig.~\ref{fig:scl90scale3pc123}  present the
principal components  and
principal planes of superclusters. 
Table~\ref{tab:pca3scale} shows that the first two principal 
components account for  99\% of the total variance of the parameters.
The highest positive values of PC3 in Fig.~\ref{fig:scl90scale3pc123}
come from high-luminosity, very elongated superclusters.
The values of PC3 for superclusters SCl~061 and SCl~094 were much 
higher than for other superclusters, therefore we excluded them
from the calculations as outliers.
These are the richest, most luminous, and most elongated systems with
the largest clumpiness in our sample
\citep[Table~\ref{tab:basicscl} and ][]{2011A&A...532A...5E}. 
The supercluster SCl~061 is
the richest member of the Sloan Great Wall, an exceptional
system in the nearby universe \citep{2011ApJ...736...51E, 2011arXiv1105.3378S}.
The supercluster SCl~094 (the Corona Borealis supercluster) is the richest
system in the dominant supercluster plane   
\citep{2011A&A...532A...5E}. This system has been studied by
\citet{1998ApJ...492...45S}; look also at the references in
\citet{2011A&A...532A...5E}.

Equation~(\ref{eq:scalinglkd}) and Fig.~\ref{fig:scl90scale}
show the resulting scaling relation. 
\begin{equation}\label{eq:scalinglkd}
\log(L_g) = (5.11 K_2 - 5.87 K_1 - 0.76)\cdot\log(D) + 1.29,
\end{equation}
where $D$ denotes diameter.
The standard deviation for the
relation $sd = 0.414$. Most of the scatter comes from the parameters
of luminous superclusters, for them $sd = 0.507$, for low-luminosity
superclusters $sd = 0.183$.

In Fig.~\ref{fig:scl90scale} we denote the high-luminosity superclusters with different
symbols, according to their shape parameter. 
Figure~\ref{fig:scl90scale} shows that more elongated and less
elongated high-luminosity superclusters populate the 
$L_{g{\mathrm{(observed)}}}$-$L_{g{\mathrm{(predicted)}}}$ plane differently.
This suggests that 
luminous  superclusters can be divided into two populations according to their shapes. 
Our calculations show that there is no such difference
for low-luminosity superclusters.
The differences between the observed and predicted luminosity are the largest
for five systems  with the
highest predicted luminosity in Fig.~\ref{fig:scl90scale}. 
These are very elongated luminous superclusters in the sample, systems
of (multibranching) filament morphology,  SCl~064, 
SCl~189, SCl~336, and SCl~474, and  a multispider SCl~530
\citep[for morphological classification of superclusters we refer to]
[]{2011A&A...532A...5E}. 

Next we derived the scaling relations separately for more elongated
and less elongated high-luminosity 
superclusters (correspondingly, Eq.~(\ref{eq:scalinglkdhf}) and
Eq.~(\ref{eq:scalinglkdhs})), 
 and for all low-luminosity superclusters (Eq.~(\ref{eq:scalinglkdl})):

\begin{equation}\label{eq:scalinglkdhf}
\log(L_g) = (0.22 K_2 - 1.67 K_1 + 1.45)\cdot\log(D) + 0.69
\end{equation}
\begin{equation}\label{eq:scalinglkdhs}
\log(L_g) = (3.45 K_2 - 3.95 K_1 + 0.50)\cdot\log(D) + 2.09
\end{equation}
\begin{equation}\label{eq:scalinglkdl}
\log(L_g) = (63.80 K_2 - 62.28 K_1 - 1.52)\cdot\log(D) + 3.81
\end{equation}

\begin{figure}[ht]
\centering
\resizebox{0.40\textwidth}{!}{\includegraphics*{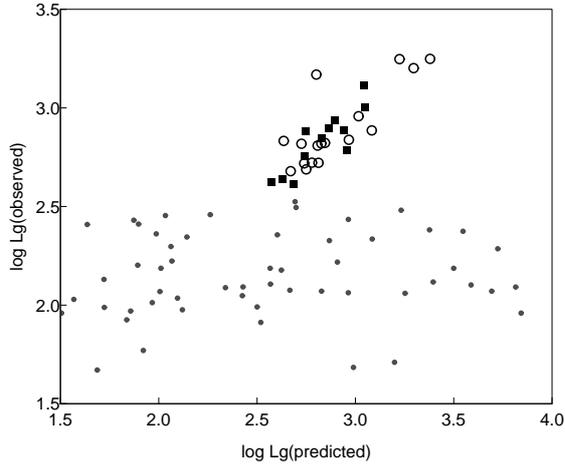}}
\caption{
$L_{g{\mathrm{(observed)}}}$ vs. $L_{g{\mathrm{(predicted)}}}$,
in units of $10^{10}h^{-2} L_{\sun}$.
Open circles denote high-luminosity superclusters with 
the luminosity $L_g >  400$ $10^{10}h^{-2} L_{\sun}$ and the
shape parameter $K_1/K_2 > 0.5$
(less elongated superclusters), squares denote
high-luminosity superclusters with 
the luminosity $L_g >  400$ $10^{10}h^{-2} L_{\sun}$ and the
shape parameter $K_1/K_2 < 0.5$
(more elongated superclusters), and
grey dots denote superclusters of lower luminosity.
}
\label{fig:scl90scalepop}
\end{figure}

\begin{figure}[ht]
\centering
\resizebox{0.22\textwidth}{!}{\includegraphics*{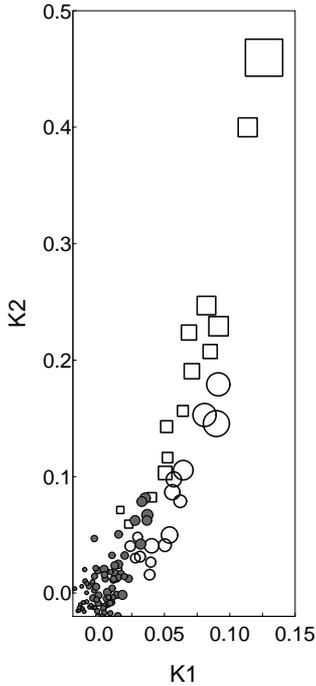}}
\caption{
Shapefinders $K_1-K_2$ plane for superclusters. 
The size of symbols is proportional to the
diameters of superclusters. 
Open circles denote high-luminosity superclusters with 
the luminosity $L_g >  400$ $10^{10}h^{-2} L_{\sun}$ and the
shape parameter $K_1/K_2 > 0.5$
(less elongated superclusters), squares denote
high-luminosity superclusters with 
the luminosity $L_g >  400$ $10^{10}h^{-2} L_{\sun}$ and the
shape parameter $K_1/K_2 < 0.5$
(more elongated superclusters),
grey filled circles denote superclusters of lower luminosity.
}
\label{fig:sclk1k2diam}
\end{figure}

Figure~\ref{fig:scl90scalepop} demonstrates
the observed vs. predicted luminosity of superclusters found
with these relations. Now luminosities of high-luminosity superclusters
are recovered well, with a very small scatter
($sd = 0.16$ and $sd = 0.22$ for more elongated and less elongated
superclusters). Interestingly, this figure shows the absence  of 
the correlation between the observed and predicted luminosity for
low-luminosity superclusters. To understand this, we plot in
Fig.~\ref{fig:sclk1k2diam} the shapefinders $K_1-K_2$ plane
for superclusters where the size of symbols is proportional to the
diameters of superclusters.
Here the values of shapefinders for 
high-luminosity superclusters are correlated, and these superclusters
 also have larger sizes. Most low-luminosity superclusters have very low,
uncorrelated 
values of shapefinders (both $K_1$ and $ K_2$ $< 0.025$).
For the smallest systems they are even negative. An example
of such a system is the Virgo supercluster \citep{e07}.
The results of the PCA show  that, 
while pairwise correlations between the luminosity and other parameters
in Table~\ref{tab:rank} are strong, the 
correlations between several parameters 
(diameters, shapefinders, and luminosities
for most of low-luminosity  superclusters)  are almost absent,
and so the scaling relation for them cannot be derived.
Correlation between the observed and predicted luminosity for low-luminosity
superclusters in Fig.~\ref{fig:scl90scale} comes from the 
high-luminosity superclusters and from the low-luminosity superclusters
with the shape parameters $K_1$ and $ K_2$ $> 0.025$.
                           
\section{Selection effects  }
\label{sect:sel}

The main selection effect in our study comes from the use of a flux-limited 
sample of galaxies to determine the luminosity density field and superclusters.
To have luminosity-dependent selection effects as small as possible, we used data 
about galaxies and galaxy systems from a distance interval
90 -- 320~\Mpc, in which these effects are the least (we refer to T10
for details). We showed above that the  parameters of
superclusters (except the number of galaxies) do not correlate
with distance, which shows that the distant-dependent 
selection effects are correctly taken into account when generating the supercluster
catalogue.

If the number of cells used to define superclusters is too small then the
supercluster catalogue may include objects that cannot be considered as real
superclusters. Moreover, the detection of the shape parameter becomes
unreliable. If the shape parameter is determined using the inertia tensor
method then superclusters have to be defined using at least eight members
\citep{2001MNRAS.320...49K}. In our study we determine shapefinders with
Minkowski functionals,  and the minimum number of cells for defining superclusters
is 64  (Appendix~\ref{sec:DF}). We analysed systems in a distance interval where
the selection effects are small. Even the poorest systems contain 
at least 25 to 30 galaxies and several
groups of galaxies. Therefore the detection of the shape parameter  
may only be affected weakly by the  selection effects 
except for the poorest systems of ``spider''  
morphology for which the shapefinders may be noisy. 
We note that \citet{2011MNRAS.411.1716C} include 
systems with at least ten member 
galaxies in their supercluster catalogue
to study of the shape parameter of superclusters.

Another selection effect comes from the choice of the threshold density to 
determine superclusters. At the density level used in the present paper 
($D = 5.0)$, rich superclusters do not percolate yet. 
If we use a lower threshold 
density, new galaxies are added to superclusters, and some 
superclusters may join to form huge systems. 
At a higher density level, galaxies in the outskirts of the
superclusters no longer belong to superclusters, and superclusters
become poorer and smaller. 

\begin{table}[ht]
\caption{Results of the principal component analysis for the threshold
density level $D = 5.5$.
}
\begin{tabular}{lrrrrr} 
\hline\hline 
(1)&(2)&(3)&(4)&(5)&(6)\\      
\hline 
            &   PC1   &   PC2   &   PC3   &  PC4 &  PC5     \\
\hline
log$N_{\mbox{gal}}$  &   -0.437 &   0.085 &   0.889 &  -0.082 &   0.052   \\
log$L_g$   &   -0.460 &   0.093 &  -0.146 &   0.854 &  -0.166  \\
log$Diameter$& -0.447 &   0.523 &  -0.282 &  -0.443 &  -0.498    \\
log$Volume$  &   -0.461 &   0.094 &  -0.298 &  -0.150 &   0.816  \\
log$D_{peak}$ &   -0.428 &  -0.837 &  -0.136 &  -0.208 &  -0.233  \\
\hline
\multicolumn{3}{l}{Importance of components} && \\ 
\hline
     &          PC1  &   PC2  &   PC3  &  PC4 &  PC5    \\
St.deviation    & 2.127 &  0.484 &  0.406 &  0.214 &  0.172  \\
Prop.Variance   & 0.904 &  0.046 &  0.033 &  0.009 &  0.005  \\
Cum.Proportion  & 0.904 &  0.951 &  0.984 &  0.994 &  1.000  \\
\hline

\label{tab:pca55}  
\end{tabular}\\
\tablefoot{ 
Notations as in Table~\ref{tab:pca90dist}.
}
\end{table}

\begin{table}[ht]
\caption{Results of the Spearman's rank correlation test
 for the threshold
density level $D = 5.5$.}
\begin{tabular}{lrr} 
\hline\hline 
(1)&(2)&(3)\\      
\hline 
Parameters     & $r$ & $p$  \\
\\
$\log(L_g)$ vs. $\log(N_{\mbox{gal}})$    & 0.85  & $< 2.2e-16$   \\
$\log(L_g)$ vs. $\log(Diameter)$                 & 0.94  & $< 2.2e-16$   \\
$\log(L_g)$ vs. $\log(Volume)$              & 0.98  & $< 2.2e-16$   \\
$\log(L_g)$ vs. $\log(D_{\mbox{peak}})$   & 0.94  & $< 2.2e-16$   \\
\hline            
                
\label{tab:rank55}  
\end{tabular}\\
\tablefoot{ 
Rank correlation coefficient  $r$ and the p-value $p$.
}
\end{table}

To see the sensitivity of the PCA results to the small differences in the choice 
of the threshold density, we compared the results of 
the PCA for superclusters chosen at higher and lower threshold density levels. As an 
example we show in Table~\ref{tab:pca55} the coefficients of the principal 
components for the superclusters chosen at the threshold density level $D = 5.5$. 
At 
this density level, \citet{2011MNRAS.415..964L} determined superclusters in the 
SDSS-DR7 for volume-limited samples of galaxies. We used flux-limited samples,
thus the density levels cannot be compared directly, but we can still choose this 
level for the present test. 
Table~\ref{tab:rank55} shows the results of the Spearman's correlation test for 
this density level. 
The comparison with Tables~\ref{tab:pca90} and ~\ref{tab:rank} shows that 
the coefficients are almost the same. 
Therefore the results of the
correlation test and  the PCA are not very sensitive to the choise of the density level.

\section{Discussion and conclusions}
\label{sect:disc}

We studied the properties
of superclusters drawn from the SDSS DR7 using the principal component analysis and
Spearman's correlation test. 
Several earlier studies have shown that 
the properties of superclusters are correlated 
(see the references in Sect.~\ref{sect:intro}).
However, it is surprising that the correlations between the various 
properties of superclusters are so tight.
The first two principal components account for  
most of the variance in the data. Different physical parameters (the luminosity, 
volume, and  diameter)  and the morphological parameters (the clumpiness 
and the shape parameters)  are almost equally important in 
shaping the properties of superclusters. 
This suggests that superclusters, as described by their overall
physical and morphological properties and by their inner morphology
and peak density, are 
objects that can be described with a few 
parameters. We derived the scaling relation for superclusters
in which we combine their luminosities, diameters, and shapefinders.

We saw  in 
Fig.~\ref{fig:scl90scale} that  more elongated and less
elongated high-luminosity superclusters populate the 
$L_{g(observed)}$-$L_{g(predicted)}$ plane differently.
This suggests that  
luminous  superclusters can be divided into two populations 
according to their shapes -- more elongated systems with the 
shape parameter $K_1/K_2 < 0.5$ and less elongated ones with
$K_1/K_2 > 0.5$. 
 \citet{2011A&A...532A...5E}  got a similar result using  multidimensional normal
mixture modelling. 
It is remarkable that two different multivariate methods
reveal information about the data in such good agreement.
However, there are few high-luminosity
superclusters in our sample. There are 14 systems with the shape
parameter $K_1/K_2 < 0.5$ among them, and 
17 systems with  $K_1/K_2 > 0.5$.
A larger sample of superclusters 
has to be analysed to confirm this result.

Parameters
used to characterise superclusters in the present study do not reflect
all the properties of superclusters. For example, rich superclusters contain 
high-density cores that  may contain merging X-ray clusters 
and may be collapsing \citep{1998ApJ...492...45S, 2000MNRAS.312..540B, e2001, rose02,
2007A&A...464..815E, e08}.
A supercluster environment with  a wide range of densities affects the properties of 
galaxies, groups, and clusters located there \citep{e2003b, 2004ogci.conf...19P, 
2005A&A...443..435W, 2006MNRAS.371...55H, e07b, 
2008MNRAS.388.1152P, tempel09, 2010ASPC..423...81F, 2011A&A...529A..53T,
2011ApJ...736...51E}. \citet{2011ApJ...736...51E} showed that the dynamical evolution of 
one of the richest superclusters in the Sloan Great Wall (SCL~111,
SCl~024 in L10 catalogue) is almost finished, while
the richest member of the Wall, SCl~126 (SCl~061) is still dynamically
active. 
Therefore our results reflect only certain aspects 
of the properties of superclusters.
 
Systems of galaxies determined in the SDSS have been studied by a number of authors
\citep{2005MNRAS.357.1068P, gott05,  park05, 2006MNRAS.372..827P, gott08, 2008MNRAS.387..767P, 
2009MNRAS.400..183K, 2010ApJS..190..181C, 
2011MNRAS.414..384S, 2011ApJ...736...51E, 2011A&A...532A...5E,
2011arXiv1105.3378S, 2011MNRAS.410.1837P, 2011MNRAS.tmp.1062P}.
The overall shapes of superclusters have been described by 
the shape parameters or approximated by triaxial ellipses 
\citep{1998A&A...336...35J, 2001MNRAS.323...47B, 
kbp02, bas03, 2007A&A...462..397E, 
2011ApJ...736...51E, 2011A&A...532A...5E, 
2011MNRAS.411.1716C, 2011MNRAS.415..964L}. These studies 
showed that elongated, prolate structures dominate among superclusters. 
The results obtained using
the moments of inertia tensor \citep{2001MNRAS.323...47B, bas03} 
or the Minkowski functionals  are in a good agreement 
\citep[see also][]{e07, 2011A&A...532A...5E}. 
In addition, \citet{2006MNRAS.365..539B} analysed correlations between supercluster
properties from simulations and find that the 
amplitude of the supercluster - cluster alignment
increases (weakly) with superclusters filamentarity.

The properties of superclusters are determined by their formation and evolution. 
\citet{kbp02}  show that the shapes of 
superclusters agree better with a $\Lambda$CDM model than with a
$\tau$CDM model. Also \citet{2011MNRAS.415..964L} found that the shapes of observed
superclusters agree with those in the $\Lambda$CDM model.
In the $\Lambda$CDM concordance cosmological model,  the matter density 
$\Omega_{\mathrm{m}}$ dominated in the early universe and the structures formed by 
hierarhical clustering driven by gravity. As the universe expands, the average 
matter density decreases. At a certain epoch, the  dark energy density 
$\Omega_{\mathrm{\Lambda}}$ became higher than the matter density,
and the universe  started to expand acceleratingly. 
Simulations of the evolution and the future of the structure in an accelerating 
universe 
show the freezing of the web -- the large-scale evolution of  structures  slows down 
\citep[][and references therein] {2002PhRvD..65d7301L, 2003NewA....8..439N, 
2006MNRAS.366..803D, 2007JCAP...10..016H, 2007GReGr..39.1545K}. 
\citet{2009MNRAS.399...97A} show that this affects the sizes, the shapes, and 
the inner structure of superclusters, and they become rounder, smaller, and their
multiplicity decreases. According to our present results, this suggests that 
in the future  superclusters become
less elongated and the scatter in the scaling relation of superclusters 
may decrease.

Summarising, our study showed that

\begin{itemize}

\item[1)] 
The PCA and Spearman's correlation test 
showed the absence of correlations between the physical properties
of superclusters and their distance, therefore the distance-dependent
selection effects were taken into account properly when generating
supercluster catalogues.
\item[2)] 
The correlations between the properties of superclusters are tight.
Different 
physical parameters (the luminosity, the volume, and the diameter)  
 and the morphological parameters (the clumpiness and the shapefinders)
 of superclusters
are equally important in shaping the properties of superclusters.
\item[3)] 
The first two principal components account for more than 90\% of the
variance of the supercluster properties and define the fundamental 
plane of superclusters.
This suggests that superclusters
can be described with a few 
physical and morphological parameters. 
We derived the scaling relation for superclusters
using data about their luminosities, diameters, and shapefinders.
\item[4)] 
Superclusters can be divided into two
populations according to their luminosity, using the luminosity limit
$L_{\mbox{g}} = 400$ $10^{10}h^{-2} L_{\sun}$.
In agreement with \citet{2011A&A...532A...5E}, we find
that high-luminosity superclusters can be divided into two sets:
more elongated systems with the 
shape parameter $K_1/K_2 < 0.5$ and less elongated ones with
$K_1/K_2 > 0.5$.
\end{itemize}

For our study we chose a small sample of superclusters least affected by 
selection effects.  To understand the properties of superclusters better the 
next step is to study  a large sample of superclusters and  high-redshift 
superclusters. 
A few superclusters at very high redshifts have already been 
discovered \citep{2005MNRAS.357.1357N, 2007MNRAS.379.1343S, 2008ApJ...684..933G, 
tanaka09, 2011arXiv1101.2024P, 2011A&A...532A..57S}. Deep surveys like the 
ALHAMBRA project \citep{moles08} will provide us with data about (possible) very 
distant superclusters.  
 We also need more simulations with various cosmologies  to 
understand the evolution and the properties of superclusters in detail.

\begin{acknowledgements}

We thank the referee, Dr. S. Basilakos, for the comments and suggestions 
that helped to improve the paper.  
Funding for the Sloan Digital Sky Survey (SDSS) and SDSS-II has been
the National Science Foundation, the U.S.  Department of Energy, the
National Aeronautics and Space Administration, the Japanese Monbukagakusho,
and the Max Planck Society, and the Higher Education Funding Council for
England.  The SDSS Web site is http://www.sdss.org/.

The SDSS is managed by the Astrophysical Research Consortium (ARC) for the
Participating Institutions.  The Participating Institutions are the American
Museum of Natural History, Astrophysical Institute Potsdam, University of
Basel, University of Cambridge, Case Western Reserve University, The
University of Chicago, Drexel University, Fermilab, the Institute for
Advanced Study, the Japan Participation Group, The Johns Hopkins University,
the Joint Institute for Nuclear Astrophysics, the Kavli Institute for
Particle Astrophysics and Cosmology, the Korean Scientist Group, the Chinese
Academy of Sciences (LAMOST), Los Alamos National Laboratory, the
Max-Planck-Institute for Astronomy (MPIA), the Max-Planck-Institute for
Astrophysics (MPA), New Mexico State University, Ohio State University,
University of Pittsburgh, University of Portsmouth, Princeton University,
the United States Naval Observatory, and the University of Washington.

We acknowledge the
Estonian Science Foundation for support under grants No.  8005 and
7146, 7765, and the Estonian Ministry for Education and Science support by grant
SF0060067s08. 
This work has also been supported by
ICRAnet through a professorship for Jaan Einasto,  by
the University of Valencia through a visiting professorship for Enn Saar and
by the Spanish MEC project AYA2006-14056, ``PAU'' (CSD2007-00060), including
FEDER contributions, and  the Generalitat Valenciana project of excellence 
PROMETEO/2009/064. 
   The density maps and the supercluster catalogues were
calculated at the High Performance Computing Centre, University of Tartu.
In this paper we use {\it R}, an open-source free statistical environment 
developed under the GNU GPL \citep[][\texttt{http://www.r-project.org}]{ig96}. 

\end{acknowledgements}

\bibliographystyle{aa}
\bibliography{pcabib.bib}
\begin{appendix}

\section{Luminosity density field and superclusters}
\label{sec:DF}

To calculate the luminosity density field, we  calculate
the luminosities of groups first. In flux-limited samples, galaxies outside
the observational window remain unobserved. To take into account
the luminosities of the galaxies that lie outside the sample limits also
 we multiply the observed galaxy
luminosities by the weight $W_d$.  The distance-dependent weight
factor $W_d$ was calculated as  
\begin{equation}
    W_d =  {\frac{\int_0^\infty L\,n
    (L)\mathrm{d}L}{\int_{L_1}^{L_2} L\,n(L)\mathrm{d}L}} ,
    \label{eq:weight}
\end{equation}
where $L_{1,2}=L_{\sun} 10^{0.4(M_{\sun}-M_{1,2})}$ are the luminosity 
limits of the observational window at a distance $d$, corresponding to the 
absolute magnitude limits of the window $M_1$ and $M_2$; we took 
$M_{\sun}=4.64$\,mag in the $r$-band \citep{2007AJ....133..734B}. 
Due to their peculiar velocities, 
the distances of galaxies are somewhat uncertain; if the galaxy belongs to a 
group, we use the group distance to determine the weight factor. 

\begin{figure}[ht]
\centering
\resizebox{0.40\textwidth}{!}{\includegraphics*{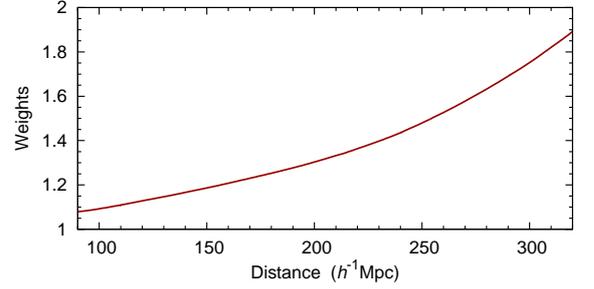}}
\caption{Weights used to correct for probable group members outside the
  observational luminosity window.}
\label{fig:weight}
\end{figure}

The luminosity weights for the groups of the SDSS DR7 in the distance interval
90~\Mpc $\ge D \le $ 320~\Mpc\
are plotted as a
function of the distance from the observer in Fig.~\ref{fig:weight}.  
The mean weight is slightly higher than unity (about 1.4)
within the sample limits.  
When the distance is greater, the weights
increase owing to the absence of faint galaxies.  
Details of the calculations of weights are given also in
 \citet{2011A&A...529A..53T}.
In the final flux-limited group catalogue, the richness of groups
decreases rapidly at distances $D > 320$~\Mpc\ due to selection
effects \citep{2010A&A...514A.102T, 2011A&A...532A...5E}.
This is another reason to choose for our study superclusters
from the distance interval 90~\Mpc\ $\le D \le $ 320~\Mpc\, where
the selection effects are weak. 
Even the poorest systems in our sample contain several groups of 
galaxies being real galaxy systems comparable to 
the Local supercluster.

To calculate a luminosity density field, 
we convert the spatial positions of galaxies $\mathbf{r}_i$ 
and their luminosities  $L_i$ into
spatial (luminosity) densities using kernel densities
\citep{silverman86}:
\begin{equation}
    \rho(\mathbf{r}) = \sum_i K\left( \mathbf{r} - \mathbf{r}_i; a\right) L_i,
\end{equation}
where the sum is over all galaxies, and $K\left(\mathbf{r};
a\right)$ is a kernel function of a width $a$. Good kernels
for calculating densities on a spatial grid are generated by box splines
$B_J$. Box splines are local and they are interpolating on a grid:
\begin{equation}
    \sum_i B_J \left(x-i \right) = 1,
\end{equation}
for any $x$ and a small number of indices that give non-zero values for $B_J(x)$.
We use the popular $B_3$ spline function:
        \begin{eqnarray}
        B_3(x)&=&\left(|x-2|^3-4|x-1|^3+6|x|^3-\right.\nonumber\\
                &&\left.-4|x+1|^3+|x+2|^3\right)/12.
        \end{eqnarray}
The (one-dimensional) $B_3$ box spline kernel $K_B^{(1)}$ of the width $a$ is defined as
\begin{equation}
    K_B^{(1)}(x;a,\delta) = B_3(x/a)(\delta / a),
\end{equation}
where $\delta$ is the grid step. This kernel differs from zero only
in the interval $x\in[-2a,2a]$. It is close to a Gaussian with $\sigma=0.6$ in the
region $x\in[-a,a]$, so its effective width is $2a$ \citep[see, e.g.,][]{saar09}.
The kernel preserves the
interpolation property exactly for all values of $a$ and $\delta$,
where the ratio $a/\delta$ is an integer. (This kernel can be used also if this ratio
is not an integer, 
and $a \gg \delta$; the kernel sums to 1 in this case, too, with a very small error.)
This means that if we apply this kernel to $N$ points on a one-dimensional grid,
the sum of the densities over the grid is exactly $N$.
 
The three-dimensional kernel $K_B^{(3)}$
is given by the direct product of three one-dimensional kernels:
\begin{equation}
    K_B^{(3)}(\mathbf{r};a,\delta) \equiv K_3^{(1)}(x;a,\delta) K_3^{(1)}(y;a,\delta) K_3^{(1)}(z;a,\delta),
\end{equation}
where $\mathbf{r} \equiv \{x,y,z\}$. Although this is a direct product,
it is isotropic to a good degree \citep{saar09}.

In \citet{e07} we compared the Epanechnikov, the Gaussian, and  $B_3$ box spline kernels
for calculating the density field. The Epanechnikov and the $B_3$ kernels
are both compact, while the Gaussian kernel is infinite and
has to be cut off at a fixed radius, which introduces an extra parameter.
We also found that both the Epanechnikov and the $B_3$ kernels
describe the overall shape of 
superclusters well, while the $B_3$ box spline kernel resolves the inner 
structure of superclusters better. This is  why we used this kernel in 
the present study.

The densities were calculated on a cartesian grid based on the SDSS $\eta$,
$\lambda$ coordinate system, as it allowed the most efficient fit of
the galaxy sample cone into a brick.
Using the rms velocity $\sigma_v$, translated into distance,
and the rms projected radius $\sigma_r$ from the group catalogue (T10),
we suppress the cluster finger redshift distortions. We
divide the radial distances between the group galaxies and the group centre by the ratio
of the rms sizes of the group finger:
\begin{equation}
    d_\mathrm{gal,f} = d_\mathrm{group} + (d_\mathrm{gal,i} - 
    d_\mathrm{group})\; \sigma_\mathrm{r} / \sigma_\mathrm{v}.
\end{equation}
This removes the smudging effect the fingers have on the density field.

The grid coordinates are calculated according to Eq.\ref{eq:xyz}.
We used an 1~\Mpc\ step grid and chose the kernel width $a=8$~\Mpc.
This kernel differs from zero within the radius 16~\Mpc,
but significantly so only inside the 8~\Mpc\ radius.
As a lower limit for the volume of superclusters we used the value
$(a/2)$~\Mpc$^3$ (64 grid cells). In this way we exclude small spurious density field objects
which include almost no galaxies.
\citet{2010arXiv1012.1989J} tested the method generating the 
superclusters from the Millenium simulations. This comparison
showed that supercluster algorithms work well, and, in addition,
  the selection effects 
have been properly taken into account when generating a supercluster
catalogue from flux-limited sample of galaxies.

Before extracting superclusters we apply the DR7 mask 
constructed by P.~Arnalte-Mur \citep{martinez09, 2010arXiv1012.1989J} 
 to the density field
and convert densities into units of mean density. The mean
density is defined as the average over all pixel values inside the mask. The mask is
designed to follow the edges of the survey and the galaxy distribution inside the
mask is assumed to be homogeneous.

\section{Minkowski functionals and shapefinders} 
\label{sec:MF}

The supercluster morphology is fully characterised by the four Minkowski 
functionals \mbox{$V_0$--$V_3$}. 
For a given surface the four Minkowski functionals (from the first to the
fourth) are proportional to the enclosed volume $V$, the area of the surface
$S$, the integrated mean curvature $C$, and the integrated Gaussian curvature
$\chi$ \citep{sah98, mar03, sss04, saar06, saar09}.

With the first three Minkowski functionals, we calculate the dimensionless 
shapefinders $K_1$ (planarity) and $K_2$ (filamentarity) \citep{sah98, sss04}. 
See also \citet{2001MNRAS.323...47B}, 
in this study the shapefinders were determined with the moments of
inertia method. 
First we 
calculate the shapefinders $H_1$--$H_3$ with a combination of Minkowski 
functionals: $H_1=3V/S$ (thickness), $H_2=S/C$ (width), and $H_3=C/4\pi$ 
(length).  Then we use the shapefinders $H_1$--$H_3$ to calculate two 
dimensionless shapefinders $K_1$ (planarity) and $K_2$ (filamentarity): $K_1 = 
(H_2 - H_1)/(H_2 + H_1)$ and $K_2 = (H_3 - H_2)/(H_3 + H_2)$. We characterise 
the overall shape of superclusters using planarity $K_1$ and filamentarity 
$K_2$, and their ratio, $K_1$/$K_2$ (the shape parameter). 

The fourth Minkowski functional $V_3$, describes the topology of the surface and 
gives the number of isolated clumps, the number of void bubbles, and the 
number of tunnels (voids open from both sides) in the region \citep[see, 
e.g.][]{saar06}. Morphologically the superclusters with low values of the 
fourth Minkowski functional $V_3$ can be described as simple spiders or simple 
filaments. High values of 
the fourth Minkowski functional $V_3$ suggest a complicated (clumpy) morphology 
of a supercluster, described as multispiders or multibranching filaments 
\citep{e07, 2011A&A...532A...5E}.

\section{Data on luminous ($L_g > 400$ $10^{10}h^{-2} L_{\sun})$ 
superclusters} 
\label{sec:Datascl}

\begin{table*}[ht]
\caption{Data on luminous ($L_g > 400$ $10^{10}h^{-2} L_{\sun})$ 
superclusters}
\begin{tabular}{rrrrrrrrrrrrr} 
\hline\hline  
(1)&(2)&(3)&(4)&(5)& (6)&(7)&(8)&(9)&(10)&(11)&(12)& (13)\\      
\hline 
 ID   &\multicolumn{1}{c}{ID}& $Distance$ & $L_{\mbox{g}}$ & $N_{\mbox{gal}}$ &  
 $Volume$& $Diameter$  & $D_{\mbox{peak}}$ & $V_3$ & $K_1$ & $K_2$ & $K_1/K_2$ & $ID_{E01}$\\
  & & Mpc/$h$ & $10^{10}h^{-2} L_{\sun}$ & & $(\vmh)^{3}$ & Mpc/$h$   &  &  &  & &  \\
\hline
1   & 239+027+009 & 264 &   1591.5  &  1038  &   8435   &  50  &  21.6  &   2  &  0.080  &  0.152  &   0.527  & 162    \\
10  & 239+016+003 & 111 &    680.2  &  1463  &   3378   &  22  &  16.4  &   1  &  0.038  &  0.015  &   2.456  & 160   \\
11  & 227+006+007 & 233 &   1476.0  &  1222  &   8065   &  35  &  16.7  &   4  &  0.053  &  0.049  &   1.081  & 154   \\
24  & 184+003+007 & 230 &   1768.2  &  1469  &   10040  &  56  &  14.1  &   5  &  0.089  &  0.145  &   0.616  & 111   \\
38  & 167+040+007 & 224 &    660.7  &   586  &   3243   &  22  &  13.8  &   2  &  0.023  &  0.040  &   0.593  &  95   \\
55  & 173+014+008 & 242 &   1773.0  &  1306  &   9684   &  50  &  12.3  &   5  &  0.091  &  0.179  &   0.509  & 111    \\
60  & 247+040+002 &  92 &    527.4  &  1335  &   2472   &  21  &  12.0  &   2  &  0.013  &  0.021  &   0.645  & 160   \\
61  & 202-001+008 & 255 &   4315.3  &  3056  &   23475  & 106  &  12.9  &  13  &  0.126  &  0.459  &   0.274  & 126 \\
64  & 250+027+010 & 301 &   1305.4  &   619  &   6058   &  55  &  12.6  &   4  &  0.091  &  0.229  &   0.399  & 164   \\
87  & 215+048+007 & 213 &    477.8  &   445  &   2301   &  21  &  11.0  &   2  &  0.039  &  0.026  &   1.494  &     \\
94  & 230+027+006 & 215 &   2263.4  &  1830  &   11256  &  54  &  11.1  &   8  &  0.113  &  0.399  &   0.284  & 158    \\
129 & 170+053+010 & 309 &    526.7  &   223  &   2321   &  20  &  10.6  &   3  &  0.029  &  0.048  &   0.612  &  \\
136 & 189+017+007 & 212 &    523.2  &   504  &   2590   &  20  &  10.9  &   2  &  0.027  &  0.030  &   0.925  & 271     \\
152 & 230+005+010 & 301 &    907.5  &   423  &   4756   &  32  &  10.8  &   3  &  0.057  &  0.097  &   0.585  & 160  \\
189 & 126+017+009 & 267 &    771.0  &   433  &   3063   &  43  &   9.5  &   4  &  0.070  &  0.190  &   0.372  &    \\
195 & 134+038+009 & 280 &    487.9  &   273  &   2200   &  23  &   9.9  &   2  &  0.031  &  0.031  &   1.004  &   \\
198 & 152-000+009 & 284 &    863.9  &   473  &   4448   &  38  &   9.7  &   4  &  0.050  &  0.103  &   0.490  &  82   \\
223 & 187+008+008 & 268 &    703.7  &   462  &   3368   &  33  &   9.3  &   3  &  0.051  &  0.142  &   0.361  & 111   \\
228 & 203+059+007 & 210 &    644.0  &   643  &   3361   &  31  &   9.5  &   2  &  0.040  &  0.040  &   0.992  & 133  \\
327 & 170+000+010 & 302 &    419.8  &   205  &   1747   &  20  &   8.5  &   2  &  0.016  &  0.071  &   0.228  &   \\
332 & 175+005+009 & 291 &    664.3  &   333  &   3128   &  27  &   8.2  &   3  &  0.062  &  0.078  &   0.788  & 106  \\
336 & 172+054+007 & 207 &   1003.6  &  1005  &   4605   &  53  &   8.7  &   5  &  0.082  &  0.246  &   0.332  & 109  \\
349 & 207+026+006 & 188 &    768.8  &   893  &   3942   &  42  &   8.8  &   4  &  0.064  &  0.105  &   0.610  & 138 \\
350 & 230+008+003 & 105 &    436.3  &   955  &   1987   &  22  &   8.0  &   2  &  0.022  &  0.059  &   0.383  & 160  \\
351 & 207+028+007 & 225 &    689.1  &   615  &   3292   &  32  &   8.7  &   4  &  0.056  &  0.086  &   0.647  & 138  \\
366 & 217+020+010 & 300 &    763.4  &   353  &   3681   &  31  &   8.1  &   4  &  0.064  &  0.156  &   0.409  & 158 \\
376 & 255+033+008 & 258 &    658.0  &   437  &   3097   &  27  &   8.6  &   4  &  0.050  &  0.041  &   1.228  & 167  \\
474 & 133+029+008 & 251 &    612.6  &   389  &   2299   &  43  &   7.6  &   4  &  0.068  &  0.223  &   0.307  &  76   \\
512 & 168+002+007 & 227 &    410.7  &   371  &   1658   &  26  &   7.5  &   3  &  0.040  &  0.082  &   0.490  &  91 \\
530 & 192+062+010 & 306 &    790.3  &   333  &   3690   &  40  &   7.5  &   4  &  0.084  &  0.207  &   0.409  &    \\
827 & 189+003+008 & 254 &    572.4  &   405  &   2238   &  30  &   6.7  &   4  &  0.052  &  0.116  &   0.450  & \\
                               
\label{tab:basicscl}  
\end{tabular}\\
\tablefoot{                                                                                 
Columns are as follows:
1: ID in L10 catalogue;
2: supercluster ID (AAA+BBB+ZZZ, AAA -- R.A., +/-BBB -- Dec., CCC -- 100$z)$;
3: the distance of the supercluster;
4: the total weighted luminosity of galaxies in the supercluster, $L_{\mbox{g}}$;
5: the number of galaxies in a supercluster, $N_{\mbox{gal}}$;
6: the volume of the supercluster, $Volume$;
7: the supercluster diameter, $Diameter$ (the maximum distance between galaxies in
the supercluster);
8: the peak density $D_{\mbox{peak}}$ of the supercluster, in units of mean density;
9: the maximum value of the fourth Minkowski functional,
$V_3$ (clumpiness), for the supercluster;
10 -- 12: shapefinders $K_1$ (planarity) and $K_2$ (filamentarity), and 
the ratio of the shapefinders $K_1/K_2$ of the full supercluster.
13: $ID_{E01}$: the supercluster ID  in  the catalogue by \citet{e2001}. SCl~160 -- the Hercules
supercluster, SCl~111 and SCl~126 -- members of the Sloan Great Wall, SCl~158 -- the Corona 
Borealis supercluster, SCl~138 -- the Bootes supercluster, SCl~336 -- the Ursa Majoris supercluster.
}
\end{table*}

\end{appendix}

\end{document}